\documentclass[aps,pra,showpacs,floatfix,nofootinbib]{article}

\usepackage{amssymb}
\usepackage{amsmath}

\newtheorem{theorem}{Theorem}

\newtheorem{definition}[theorem]{Definition}
\newtheorem{example}[theorem]{Example}

\newtheorem{lemma}[theorem]{Lemma}

\newtheorem{proposition}[theorem]{Proposition}
\newtheorem{remark}[theorem]{Remark}

\newenvironment{proof}[1][Proof]{\noindent\textbf{#1.} }{\ \rule{0.5em}{0.5em}}
\newdimen\dummy
\dummy=\oddsidemargin
\input{tcilatex}

\begin{document}

\title{A constructive algorithm for the Cartan decomposition of
  $SU(2^{N})$\\ 
\vspace{1cm}
\large{ Henrique N. S\'{a} Earp$^1$ and Jiannis K. Pachos$^2$}\\
\small{\it $^1$Department of Mathematics, Imperial College London, Prince
Consort Road, London, SW7 2BW, UK } \\
\small{\it $^2$Department of Applied Mathematics and Theoretical
    Physics, University of Cambridge,
Wilberforce Road, Cambridge CB3 0WA, UK}}

\maketitle

\begin{abstract}

We present an explicit numerical method to obtain the Cartan-Khaneja-Glaser
decomposition of a general element $G \in SU(2^{N})$ in terms of its
`Cartan' and `non-Cartan' components. This effectively factors $G$ in terms
of group elements that belong in $SU(2^{n})$ with $n<N$, a procedure that
an be iterated down to $n=2$. We show that every step reduces
to solving the zeros of a matrix polynomial, obtained by truncation of the
Baker-Campbell-Hausdorff formula, numerically. All computational tasks involved
are straightforward and the overall truncation errors are well under
control.

\end{abstract}

\bigskip

\section{Introduction}

In their seminal paper \cite{Khaneja-Glaser}, Khaneja and Glaser presented a
generic method to decompose `large' unitary elements in terms of `smaller'
ones. The initial unitaries can be viewed as evolution operators of a
multipartite system of spin-1/2's or as quantum algorithms acting on qubits.
Experimentally it is only possible to control the interactions between a
small number of subsystems at a time. Hence, this decomposition is of
particular interest. It allows us to address questions such as how to
optimize a dynamical evolution in terms of control procedures applied to a
small number of spins or how a certain quantum algorithm can be obtained
with the smallest possible number of experimentally feasible one and two
qubit gates.

In particular, Khaneja and Glaser show that any element of the Lie group
$\mathbf{G}=SU( 2^{N}) $ is, up to local unitaries in $SU(2)
^{\otimes N}$, determined by components generated from certain Abelian subalgebras
$\mathfrak{h}_{n}$ and $\mathfrak{f}_{n}$, $n=2,...,N$, of the Lie algebra
$\mathfrak{ su} (2^{N}) $. This is achieved by employing iteratively the
Cartan decomposition
\begin{equation*}
G=K^{\prime }HK^{\prime \prime },
\end{equation*}
where $H$ is generated by $\mathfrak{h}_{n}$ ($\mathfrak{f}_{n}$) and
the factors $K^{\prime }$ and $K^{\prime \prime }$ belong to the subgroup
$\mathbf{K}\subset \mathbf{G}$ generated by a particular subalgebra
orthogonal to $\mathfrak{h}_{n}$ ($\mathfrak{f}_{n}$). These relevant
substructures are specified in terms of a fortunate choice of basis for
$\mathfrak{su}( 2^{n}) $ that can be obtained recurrently for $n=2,...,N$.

The beauty of this result and its promising applications in quantum
algorithms \cite{Vatan,Shende}, control theory, quantum error correction
\cite{Steane} or the quantification of entanglement in multi-partite
systems \cite{Osterloh-Siewert} have motivated the search for a constructive
method to perform the decompositions for any given matrix $G\in SU( 2^{N})
$. Although substantial work has been done on the first non-trivial instance
$SU( 4) $ \cite{Kraus,SU(4)VidalCirac}, little seems to be known so far
for the higher $N$ case \cite{Bullock}.

Here we address the decomposition problem for the general group $SU(2^{N})$.
Employing a convenient truncation of the Baker-Campbell-Hausdorff (BCH)
formula, we show that the problem allows for a numerical algorithm to
calculate all such $\mathbf{KHK}$ decompositions with controlled error.
Hence, we can perform the full Khaneja-Glaser decomposition of a general
element $G\in SU( 2^{N}) $ with arbitrary computational precision.

This article is organized as follows. In Section \ref{Background} we briefly
review the  Khaneja-Glaser decomposition \cite{Khaneja-Glaser} and establish
the formalism for our approach. In Section \ref{Algorithm} we reinterpret
the problem in terms of the BCH formula; we explain how a truncation of the
BCH series renders our problem solvable by straightforward numerical tasks.
In Section \ref{Step} we give a user-friendly summary of the essential steps
involved and we finally conclude in Section \ref{Conclusions}.

\section{The Khaneja-Glaser decomposition}
\label{Background}

We consider the compact semi-simple Lie group $\mathbf{G}=SU(2^{N})$ and a
particular compact closed subgroup $\mathbf{K}\subset \mathbf{G}$; we denote
by $\mathfrak{g=su(}2^{N}\mathfrak{)}$ the Lie algebra of $\mathbf{G}$ and
by $\mathfrak{k\subset g}$ the Lie algebra of $\mathbf{K}$ understood as a
subalgebra of $\mathfrak{g}$. Since $\mathbf{G}$ is semi-simple, the Killing
form $\left\langle .,.\right\rangle $ is non-degenerate and defines a
bi-invariant metric on $\mathbf{G}$. Hence, we can define
$\mathfrak{m=k}^{\bot }$ to be the orthogonal complement of $\mathfrak{k}$
with respect to the metric. Notice that, in general, the vector space
$\mathfrak{m\subset g}$ is \emph{not} a subalgebra. Since $\mathfrak{m}$ is
determined by $\mathfrak{k}$ and the Killing metric, we shall refer to this
structure as the \emph{Lie algebra pair} $( \mathfrak{g,k})$.

We shall adopt the following typographic conventions in most cases
\begin{equation*}
\begin{tabular}{lll}
$\mathbf{G}$ & capital bold & group or subgroup \\
$G$ & capital & element of $\mathbf{G}$ \\
$\mathfrak{g}$ & German fraktur & Lie algebra or subspace \\
$g$ & normal & element of $\mathfrak{g}$
\end{tabular}
\end{equation*}

The only exception to the choice of fonts just stated will be the familiar
Pauli matrices, seen as elements of $\mathfrak{su}\left( 2\right) $, which
will be denoted by majuscules
\begin{equation*}
X=\frac{\mathbf{i}}{2}\left(
\begin{array}{cc}
0 & 1 \\
1 & 0
\end{array}
\right) \text{\quad }Y=\frac{\mathbf{i}}{2}\left(
\begin{array}{cc}
0 & -\mathbf{i} \\
\mathbf{i} & 0
\end{array}
\right) \text{\quad }Z=\frac{\mathbf{i}}{2}\left(
\begin{array}{cc}
1 & 0 \\
0 & -1
\end{array}
\right)
\end{equation*}
and the $2\times 2$ identity matrix $I=\left(
\begin{array}{cc}
1 & 0 \\
0 & 1%
\end{array}%
\right) $. Finally, we shall use an abbreviated normalized notation for their
tensor products given by
\begin{equation*}
A_{1}A_{2}...B_{j}...A_{k}\doteq \left( \frac{2}{\mathbf{i}}\right)
^{k-1}A_{1}\otimes A_{2}\otimes ...\otimes B_{j}\otimes ...\otimes A_{k},%
\text{\quad }\left.
\begin{array}{l}
A_{i}=X,Y,Z \\
B_{j}=I%
\end{array}%
\right. .
\end{equation*}%
For example, $YXI=\frac{\mathbf{i}}{2}\left(
\begin{array}{cc}
0 & -\mathbf{i} \\
\mathbf{i} & 0%
\end{array}%
\right) \otimes \left(
\begin{array}{cc}
0 & 1 \\
1 & 0%
\end{array}%
\right) \otimes \left(
\begin{array}{cc}
1 & 0 \\
0 & 1%
\end{array}%
\right) .$

\subsection{Cartan decomposition}

\begin{definition}
A \emph{Cartan decomposition} of $\mathfrak{g}$ is an orthogonal split of $
\mathfrak{g}$
\begin{equation*}
\mathfrak{g=k\oplus m}
\end{equation*}
given by a Lie algebra pair $\left( \mathfrak{g},\mathfrak{k}\right) $
satisfying the commutation relations
\begin{equation}
\begin{tabular}{lll}
$ \left[ \mathfrak{k,k}\right] \subset \mathfrak{k},$ & $\left[
\mathfrak{m}, \mathfrak{k}\right] \subset \mathfrak{m},$ & $\left[
\mathfrak{m},\mathfrak{m}\right] \subset \mathfrak{k.}$
\end{tabular}
\end{equation}
\label{Cartan}
In this case $\left( \mathfrak{g},\mathfrak{k}\right) $ is
called a \emph{ symmetric Lie algebra pair}.
\end{definition}

\begin{remark}
\label{+1-1 eigenspaces}The apparently artificial conditions in the above
definition have an interpretation in Riemannian geometry: $\mathfrak{%
g=k\oplus m}$ is a Cartan decomposition if and only if the quotient manifold
$\mathbf{G}/\mathbf{K}=\exp \left( \mathfrak{m}\right) $ is a \emph{globally
Riemannian symmetric space }\cite{KobNomi}. Such a space possesses a
canonical global \emph{involution} (i.e., an automorphism $s$ of the space
onto itself such that $s^{2}=I$) which induces naturally a linear
involution $s_{\ast }$ on $\mathfrak{g}$ that preserves the Lie algebra
structure $\left[ .,.\right] $. Since $s_{\ast }$ squares to identity, its
eigenvalues must be $\pm 1$, and the subspaces $\mathfrak{k}$ and $\mathfrak{%
m}$ emerge respectively as the $+1$ and $-1$ - eigenspaces of $s_{\ast }$.
\end{remark}

We can start exploring the Cartan decomposition $\mathfrak{ g=k\oplus m}$ by
noticing that if $\mathfrak{h\subset m}$ is a subalgebra of $\mathfrak{g}$
then, by (\ref{Cartan}), $\mathfrak{h}$ is automatically Abelian. Since
$\mathfrak{m}$ itself is not in general a subalgebra of $\mathfrak{g}$, it
is natural to look for a \emph{ maximal} (Abelian) subalgebra
$\mathfrak{h\subset m}$.

\begin{definition}
A \emph{Cartan subalgebra} of the pair $\left( \mathfrak{g},\mathfrak{k}
\right) $ is a maximal (in $\mathfrak{m}$) Abelian subalgebra $\mathfrak{
h\subset m}$.
\end{definition}

From now on we shall assume that the Cartan subalgebra $\mathfrak{h}$ refers to
the pair $\left( \mathfrak{g},\mathfrak{k}\right) $ unless stated otherwise.
The following \emph{Proposition} shows that the whole $\mathfrak{m}$ is
obtained from $\mathfrak{h}$ by the adjoint action of $\mathbf{K}$ and that
there is only one such $\mathfrak{h}$ up to this action. In the context of
the particular application we have in mind, this means $\mathfrak{h}$
carries the `essential' information about $\mathfrak{m}$.

\begin{proposition}
\label{Prop Cartan subalgebras}Let $\mathfrak{h}$ and $\mathfrak{h}^{\prime
} $ be two Cartan subalgebras; then

\begin{enumerate}
\item $\mathfrak{m=}\dbigcup
\limits_{K\in \mathbf{K}}\limfunc{Ad}_{K}\left(
\mathfrak{h}\right) $;

\item $\mathfrak{h}^{\prime }=\limfunc{Ad}_{K}\left( \mathfrak{h}\right) $
for some element $K\in \mathbf{K}$.
\end{enumerate}

\begin{proof}
Please see \emph{Appendix \ref{App Cartan thm}}.
\end{proof}
\end{proposition}

Denote $\mathbf{H}=\exp (\mathfrak{h})\subset \mathbf{G}$ the subgroup
generated by $\mathfrak{h}$. The Cartan decomposition theorem states that
any group element $G\in \mathbf{G}$ can be written as an element $H\in
\mathbf{H}$ together with left- and right-multiplications by elements
of  $\mathbf{K}$:

\begin{theorem}[Cartan decomposition]
\label{Thm Cartan decomp}The Cartan decomposition $\mathfrak{g=k\oplus m}$
induces a \emph{Cartan decomposition} of the group $\mathbf{G}$,
\begin{equation}
\mathbf{G}=\mathbf{KHK},
\end{equation}
\label{G=KHK}
where $\mathbf{H}=\exp (\mathfrak{h})$.

The Cartan decomposition of a given element $G\in \mathbf{G}$ has the form
\begin{eqnarray}
G=K_{0}\underset{M}{\underbrace{K_{1}HK_{1}^{\dag }}}=K_{0}M,
\label{G=K0M}
\end{eqnarray}
where $K_{0},K_{1}\in \mathbf{K}$, $H\in \mathbf{H}$ and $
M=K_{1}HK_{1}^{\dag }\in \exp \left( \mathfrak{m}\right) $.

\begin{proof}
Since $\mathbf{G}/\mathbf{K}=\exp \left( \mathfrak{m}\right) $, there exist
$K_{0}\in \mathbf{K}$ and $M\in \exp \left( \mathfrak{m}\right) $ such that
$G=K_{0}M.$ Let $m=\log \left( M\right) \in \mathfrak{m}$; from \emph{
Proposition \ref{Prop Cartan subalgebras}}, item $1$, there exists $
K_{1}\in \mathbf{K}$ such that $\limfunc{Ad}_{K_{1}}m=h\in \mathfrak{h},$ so
\begin{equation*}
G=K_{0}\exp \left( m\right) =K_{0}\exp \left( \limfunc{Ad}
\nolimits_{K_{1}^{\dag }}h\right) =K_{0}K_{1}\underset{H}{\underbrace{\exp
\left( h\right) }}K_{1}^{\dag }.
\end{equation*}
\end{proof}
\end{theorem}

\subsection{The Khaneja-Glaser basis}

The Khaneja-Glaser basis \cite{Khaneja-Glaser} for arbitrary $\mathfrak{su}
\left( 2^{n}\right) $ makes explicit all the structures which concern us. In
particular, the splitting $\mathfrak{g=k\oplus m}$ and the Cartan
subalgebras $\mathfrak{h}_{n}$ and $\mathfrak{f}_{n}$ are manifest.

We start with a familiar example:

\begin{example}
\label{SU(4)}For $n=2$, we propose basis elements for the Lie algebra
$\mathfrak{su}\left( 4\right) $ and organize them into subspaces
$\mathfrak{m}_{2}$ and $\mathfrak{k}_{2}$ as follows:

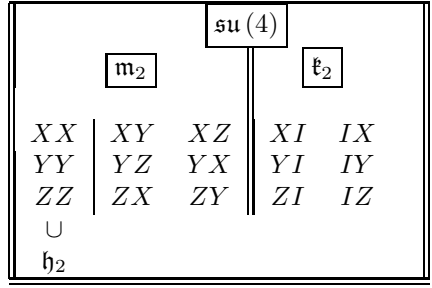
\begin{equation*}
\begin{tabular}{||cccccc||}
\hline\hline
&  & \multicolumn{2}{c}{\fbox{$\mathfrak{su}\left( 4\right) $}} &  &  \\
\multicolumn{3}{||c}{\fbox{$\mathfrak{m}_{2}$}} &
\multicolumn{2}{||c}{\fbox{$\mathfrak{k}_{2}$}} &  \\
&  &  & \multicolumn{1}{||c}{} &  &  \\
$XX$ & \multicolumn{1}{|c}{$XY$} & $XZ$ & \multicolumn{1}{||c}{$XI$} & $IX$
&  \\
$YY$ & \multicolumn{1}{|c}{$YZ$} & $YX$ & \multicolumn{1}{||c}{$YI$} & $IY$
&  \\
$ZZ$ & \multicolumn{1}{|c}{$ZX$} & $ZY$ & \multicolumn{1}{||c}{$ZI$} & $IZ$
&  \\
$\cup $ &  &  & & &  \\
$\mathfrak{h}_{2}$ &  &  &  &  &  \\ \hline\hline
\end{tabular}
\end{equation*}
\begin{eqnarray}
\mathfrak{h}_{2}&=&\limfunc{span}\left\{ XX,YY,ZZ\right\} 
\nonumber\\ \nonumber
\mathfrak{f}_{2}&=&\left\{ 0\right\}
\end{eqnarray}
\end{example}

Now we obtain the \emph{Khaneja-Glaser basis} for $\mathfrak{su}\left(
2^{n}\right) $ by a relatively simple iteration process starting from
$\mathfrak{su}\left( 4\right) $, as summarized in the 
following diagram:

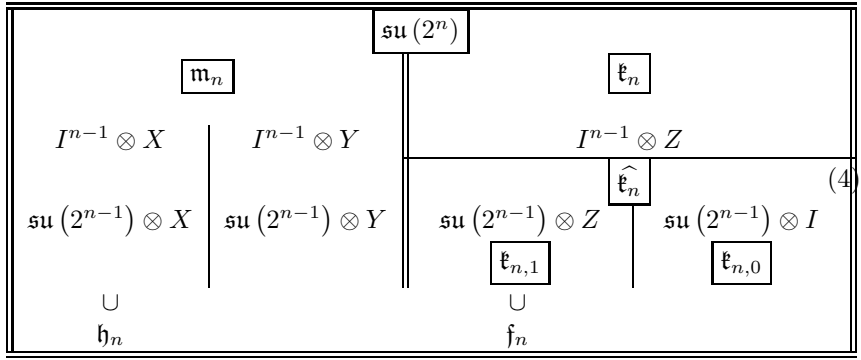
\begin{eqnarray}
&&
\begin{tabular}{||cccc||}
\hline\hline
& \multicolumn{2}{c}{\fbox{$\mathfrak{su}\left( 2^{n}\right) $}} &  \\
\multicolumn{2}{||c}{\fbox{$\mathfrak{m}_{n}$}} & \multicolumn{2}{||c||}{
\fbox{$\mathfrak{k}_{n}$}} \\
&  & \multicolumn{1}{||c}{} &  \\
$I^{n-1}\otimes X$ & \multicolumn{1}{|c}{$I^{n-1}\otimes Y$} &
\multicolumn{2}{||c||}{$I^{n-1}\otimes Z$} \\ \cline{3-4} &
\multicolumn{1}{|c}{} & \multicolumn{2}{||c||}{\fbox{$\widehat{\mathfrak{k}
_{n}}$}} \\
$\mathfrak{su}\left( 2^{n-1}\right) \otimes X$ & \multicolumn{1}{|c}{$
\mathfrak{su}\left( 2^{n-1}\right) \otimes Y$} & \multicolumn{1}{||c}{$
\underset{\fbox{$\mathfrak{k}_{n,1}$}}{
\begin{tabular}{l}
$\mathfrak{su}\left( 2^{n-1}\right) \otimes Z$
\end{tabular}
}$} & \multicolumn{1}{|c||}{$\underset{\fbox{$\mathfrak{k}_{n,0}$}}{
\begin{tabular}{l}
$\mathfrak{su}\left( 2^{n-1}\right) \otimes I$
\end{tabular}
}$} \\
$\cup $ &  & $\cup $ &  \\
$\mathfrak{h}_{n}$ &  & $\mathfrak{f}_{n}$ &  \\ \hline\hline
\end{tabular}
\label{Khaneja basis}
\end{eqnarray}
\begin{equation*}
I^{n-1}=I^{\otimes \left( n-1\right) }=\text{ }\underset{n-1}{
\underleftrightarrow{I\otimes ..\otimes I}}  \notag
\end{equation*}
NB.: Terms of the form $\mathfrak{su}\left( 2^{n-1}\right) \otimes A$ denote the set
obtained by tensoring each element of $\mathfrak{su}\left(
2^{n-1}\right)$ with the matrix $A=X,Y,X,I$;
in all cases we consider the span of the basis elements over $\mathbb{R}$.

Notice that for $n>2$ two successive Cartan decompositions can be performed.
First, the expected one referring to the pair $\left( \mathfrak{su}\left(
2^{n}\right) ,\mathfrak{k}_{n}\right) $, with Cartan subalgebra
$\mathfrak{h} _{n}$. In the terms of \emph{Theorem \ref{Thm Cartan decomp}},
this means we can write $G\in SU\left( 2^{n}\right) $ as $G=K^{\prime
}HK^{\prime \prime }$ with $H\in \exp \left( \mathfrak{h}_{n}\right) $ and
$K^{\prime},K^{\prime \prime }\in \exp \left( \mathfrak{k}_{n}\right) $.
However, the decomposition of $\mathfrak{k}_{n}$ given by diagram $
\left( \ref{Khaneja basis}\right) $ is
\begin{equation*}
\mathfrak{k}_{n}=\mathfrak{k}_{n,1}\oplus \mathfrak{k}_{n,0}\oplus \limfunc{
span}\left\{ I^{n-1}Z\right\} \simeq \mathfrak{su}\left( 2^{n-1}\right)
\oplus \mathfrak{su}\left( 2^{n-1}\right) \oplus \mathfrak{u}\left(
1\right),
\end{equation*}
where both $\mathfrak{k}_{n,1}$ and $\mathfrak{k}_{n,0}$ are canonically
isomorphic to $\mathfrak{su}\left( 2^{n-1}\right)$. Since our aim is to
iteratively decompose the factors $K^{\left( j\right) }$ until they
eventually reduce to `local' unitaries in $SU\left( 2\right) ^{\otimes N}$
and non-local `Cartan' factors, we would expect $\mathfrak{k}_{n}$ to
generate something of the form $SU\left( 2^{n-1}\right) \otimes
SU\left(2\right) $. Thus, there is a $\mathfrak{su}\left(
2^{n-1}\right)$ component
too many in $\mathfrak{k}_{n}$ that we need to factor away in order to
define the complete recurrence step.

A closer look at $\mathfrak{k}_{n}$ reveals another Lie Algebra pair, thereby
clearing the way for a second Cartan decomposition: we just have to leave
aside the `complex phase' generated by $I^{n-1}Z$ (see diagram (\ref{Khaneja
basis})) that can be seen as a `local' transformation under the inclusion
$U(1)
\hookrightarrow SU\left( 2\right) $.
Let $\widehat{\mathfrak{k}_{n} }= \mathfrak{k}_{n,1}
\oplus \mathfrak{k}_{n,0}$ denote the
subalgebra obtained from $\mathfrak{k}_{n}$ in this manner, so that
$\mathfrak{k}_{n}= \widehat{\mathfrak{k}_{n}}\oplus \mathfrak{u}\left(
1\right) \hookrightarrow \widehat{\mathfrak{k}_{n}}\oplus
\mathfrak{su}\left( 2\right) $. Accordingly, given a group element
$K=\exp \left( k\right) \in\mathbf{K}$, let us write $\widehat{K}
=\exp \left( \widehat{k}\right) $, 
where $\widehat{k}\in $ $\widehat{ \mathfrak{k}_{n}}$ is obtained from $k\in
\mathfrak{k}$ by eliminating the component
spanned by $I^{n-1}Z$. This is well defined as $I^{n-1}Z$ commutes with
every element in $\mathfrak{k}_{n}$.

It is now straightforward to check that $\left(
\widehat{\mathfrak{k}_{n}},\mathfrak{k}_{n,0}\right) $ is also a Lie algebra
pair \cite{Khaneja-Glaser}, whose Cartan subalgebra we call
$\mathfrak{f}_{n}$. Hence, we can apply 
\emph{Theorem \ref{Thm Cartan decomp}} again to decompose the factors $\widehat{K^{\prime 
}}$ and 
$\widehat{K^{\prime \prime }}$ into elements of $\exp \left
(\mathfrak{f}_{n}\right)$ together with left and right multiplication by
some new factors generated by $\mathfrak{k}_{n,0}$. This time the orthogonal
subalgebra $\mathfrak{k}_{n,0}= \mathfrak{su}\left( 2^{n-1}\right) \otimes
I$ is canonically isomorphic to $\mathfrak{su}\left( 2^{n-1}\right) $, so it
generates $\exp \left (\mathfrak{k}_{n,0} \oplus \limfunc{ span}\left\{
I^{n-1}Z\right\}\right) \simeq SU(2^{n-1}) \otimes SU(2)$. Thus, we have
accomplished the the complete $n$th recurrence step that yields the decomposition
\begin{equation*}
G=K^{\left( 1\right) }F^{\left( 1\right) }K^{\left( 2\right) }HK^{\left(
3\right) }F^{\left( 2\right) }K^{\left( 4\right) },
\end{equation*}
with $F^{\left( j\right) }\in \exp \left( \mathfrak{f}_{n}\right) ,$ $H\in
\exp \left( \mathfrak{h}_{n}\right) $ and $K^{\left( j\right) }\in SU\left(
2^{n-1}\right) \otimes SU\left( 2\right) $.

Note, finally, that we are particularly interested in the `Cartan' factors,
i.e. those generated by the Cartan subalgebras $ \mathfrak{h}_{n}$ of
$\left(\mathfrak{su}\left( 2^{n}\right) ,\mathfrak{k} _{n}\right) $ and
$\mathfrak{f}_{n}$ of $\left( \widehat{\mathfrak{k}_{n}},
\mathfrak{k}_{n,0}\right) $, that emerge in each step. It is thus
convenient to know explicitly a set of basis elements
for each of these subalgebras. This can be achieved by the following
recurrence formula, starting from
$\mathfrak{h}_{2}=\limfunc{span}\left\{XX,YY,ZZ\right\} $, 

\begin{eqnarray}
&&
\begin{tabular}{l}
$\mathfrak{h}_{n}=\limfunc{span}\mathfrak{a}\left( n\right) $ \\
$\mathfrak{f}_{n}=\limfunc{span}\mathfrak{b}\left( n\right) $%
\end{tabular}%
,\text{ }n=2,...,N  \label{recu subalgebras} \\
&&\fbox{%
\begin{tabular}{l}
$\mathfrak{a}\left( 2\right) =\left\{ XX,YY,ZZ\right\}, \mathfrak{b}\left( 2\right) =\left\{ 0\right\}  $ \\
$\mathfrak{s}\left( n\right) =\dbigcup\limits_{j=2}^{n}\mathfrak{a}\left(
j\right) \otimes I^{n-j}$ \\
$\mathfrak{a}\left( n+1\right) =\left\{ I^{n},\mathfrak{s}\left( n\right)
\right\} \otimes X$ \\
$\mathfrak{b}\left( n+1\right) =\left\{ \mathfrak{s}\left( n\right) \right\}
\otimes Z$%
\end{tabular}%
}  \notag
\end{eqnarray}

The \emph{Example} below illustrates all the above constructions for the
first nontrivial case $\mathfrak{su}\left( 8\right) $:

\begin{example}
$n=3:\mathfrak{su}\left( 8\right) $%
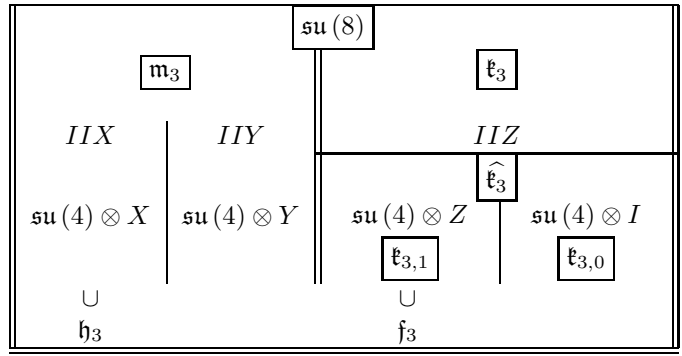
\begin{equation*}
\begin{tabular}{||cccc||}
\hline\hline
& \multicolumn{2}{c}{\fbox{$\mathfrak{su}\left( 8\right) $}} &  \\
\multicolumn{2}{||c}{\fbox{$\mathfrak{m}_{3}$}} &
\multicolumn{2}{||c||}{%
\fbox{$\mathfrak{k}_{3}$}} \\
&  & \multicolumn{1}{||c}{} &  \\
$IIX$ & \multicolumn{1}{|c}{$IIY$} & \multicolumn{2}{||c||}{$IIZ$} \\
\cline{3-4}
& \multicolumn{1}{|c}{} & \multicolumn{2}{||c||}{\fbox{$\widehat{\mathfrak{k}%
_{3}}$}} \\
$\mathfrak{su}\left( 4\right) \otimes X$ & \multicolumn{1}{|c}{$\mathfrak{su}%
\left( 4\right) \otimes Y$} & \multicolumn{1}{||c}{$\underset{\fbox{$%
\mathfrak{k}_{3,1}$}}{%
\begin{tabular}{l}
$\mathfrak{su}\left( 4\right) \otimes Z$%
\end{tabular}%
}$} & \multicolumn{1}{|c||}{$\underset{\fbox{$\mathfrak{k}_{3,0}$}}{%
\begin{tabular}{l}
$\mathfrak{su}\left( 4\right) \otimes I$%
\end{tabular}%
}$} \\
$\cup $ &  & $\cup $ &  \\
$\mathfrak{h}_{3}$ &  & $\mathfrak{f}_{3}$ &  \\ \hline\hline
\end{tabular}%
\end{equation*}

\begin{eqnarray*}
\mathfrak{h}_{3} &=&\limfunc{span}\left\{ IIX,XXX,YYX,ZZX\right\}  \\
\mathfrak{f}_{3} &=&\limfunc{span}\left\{ XXZ,YYZ,ZZZ\right\}
\end{eqnarray*}
\end{example}

\subsection{The Baker-Campbell-Hausdorff formula}

The matrix Lie algebra $\mathfrak{g}$ is noncommutative and thus, for
general elements $a,b\in \mathfrak{g}$, the product of exponentials
$e^{a}e^{b}$ does not coincide with the exponential of their sum, $e^{a+b}$.
In fact the expression for $\log \left( e^{a}e^{b}\right) $ has an infinite
series of correction terms and is known as the \emph{BCH formula},
after Baker-Campbell-Hausdorff.

Although the original formula was rather complicated and computationally
unpractical, a remarkable simplification made by Dynkin \cite{Dynkin,bose}
expresses all the terms as successive commutators of $a$ and $b$
\begin{equation*}
\log \left( e^{a}e^{b}\right) =\sum_{i,j=1}^{\infty }T_{i,j}\left(
a,b\right) .
\end{equation*}
Here $T_{i,j}\left( a,b\right) $ denotes the homogeneous term of degree $i$
in $a$ and degree $j$ in $b$; its expression is
\begin{equation*}
T_{i,j}\left( a,b\right) =\frac{1}{i+j}\sum_{\left(
i_{1},j_{1},...,i_{k},j_{k}\right) }\frac{\left( -1\right) ^{k-1}}{k}\frac{1
}{i_{1}!j_{1}!...i_{k}!j_{k}!}\left[ a^{i_{1}}b^{j_{1}}...a^{i_{k}}b^{j_{k}}
\right] ,
\end{equation*}
where we abbreviate $\left[ a^{i_{1}}b^{j_{1}}...a^{i_{k}}b^{j_{k}}\right] =
\underset{i_{1}}{\underbrace{[a,...[a,}}\underset{j_{1}}{\underbrace{[b,...[b,}}
[a,...[b,\underset{i_{k}}{\underbrace{[a,...[a,}}\underset{j_{k}}{\underbrace{
[b,...,b]...]}}$ and the sum ranges over all possible $2k$-uples of non-negative
integers $\left( i_{1},j_{1},...,i_{k},j_{k}\right) $ such that
\begin{equation*}
\sum_{c=1}^{k}i_{c}=i,\text{ }\sum_{c=1}^{k}j_{c}=j\text{ and }i_{c}+j_{c}>0.
\end{equation*}

The first few terms are
\begin{eqnarray}
\log \left( e^{a}e^{b}\right) =&&a+b+\tfrac{1}{2}\left[ a,b\right] +\tfrac{1}{
12}\left[ a,\left[ a,b\right] \right] +\tfrac{1}{12}\left[ b,\left[ b,a
\right] \right] + \nonumber \\
&&
+\frac{1}{24}\left[ a,\left[ b,\left[ a,b\right] \right] \right] +\tfrac{1}{
120}(...)\label{BCH}
\end{eqnarray}
and the higher order coefficients after $\frac{1}{120}$ decrease quickly
(see e.g. \cite{BCHcoeffs}). This will motivate us later on to perform
convenient truncations on this convergent series.

\section{Numerical algorithm for the KHK decomposition}
\label{Algorithm}

In this Section we will develop a technique that allows the explicit
numerical calculation of the components of a general group element $G\in
\mathbf{G}$ under Cartan decomposition. The idea is to consider the Cartan
decomposition (\ref{G=K0M}) in the light of the BCH expansion (\ref{BCH}).
Let $g\in \mathfrak{g}$, $m\in \mathfrak{m}$ and $k\in \mathfrak{k}$ be
the generators of $G$, $M$ and $K_{0}$, respectively. Then (\ref{G=K0M})
reads
\begin{equation}
G=e^{g}=e^{k}e^{m}.
\end{equation}
\label{eg=ekem} Since the matrix $G$ is given, $\left( \ref{eg=ekem}\right)
$ shows that $k$ can be obtained from $m$ (and vice-versa) by
\begin{equation}
k=k\left( m\right) =\log (Ge^{-m}).
\end{equation}
\label{k=log(Ge-m)} Hence, the decomposition problem is reduced to finding
$m$.

\subsection{\protect\bigskip Determining $m$}
\label{m}

First, taking logarithms in $\left( \ref{eg=ekem}\right) $, we obtain
\begin{equation}
g=\log (e^{k}e^{m}).
\end{equation}
\label{g=log(ekem)}

We then apply $\left( \ref{BCH}\right) $ to expand $g$ in terms of
successive brackets of $k$ and $m$. In the light of
\emph{Remark \ref{+1-1 eigenspaces}} we can easily deduce that
each of the brackets belongs in either $\mathfrak{k}$ or
$\mathfrak{m}$. Hence, the expansion is split into two orthogonal
components,
\begin{eqnarray*}
g &=&k+m+\underset{\in \mathfrak{m}}{\underbrace{\frac{1}{2}\left[ k,m\right]
}}+\underset{\in \mathfrak{m}}{\underbrace{\frac{1}{12}\left[ k,\left[ k,m%
\right] \right] }}+\underset{\in \mathfrak{k}}{\underbrace{\frac{1}{12}\left[
m,\left[ m,k\right] \right] }}+\underset{\in \mathfrak{k}}{\underbrace{\frac{
1}{24}\left[ k,\left[ m,\left[ k,m\right] \right] \right] }}+... \\
&=&g_{\mathfrak{k}}+g_{\mathfrak{m}},
\end{eqnarray*}
where
\begin{eqnarray}
g_{\mathfrak{k}} &=&k+\frac{1}{12}\left[ m,\left[ m,k\right] \right] +\frac{1
}{24}\left[ k,\left[ m,\left[ k,m\right] \right] \right] +...\in \mathfrak{k}
,  \label{gkgm} \\
g_{\mathfrak{m}} &=&m+\frac{1}{2}\left[ k,m\right] +\frac{1}{12}\left[ k,
\left[ k,m\right] \right] +...\in \mathfrak{m.}  \label{gkgmb}
\end{eqnarray}
Note that computing $g_{\mathfrak{k}}$ and $g_{\mathfrak{m}}$ from $g$ is a
straightforward task since the Khaneja-Glaser basis $\left( \ref{Khaneja basis}
\right) $ makes explicit the partition $\mathfrak{g=k\oplus m}$.

At this stage we can use (\ref{k=log(Ge-m)}) to eliminate $k=k\left(m\right)
$ in either of the equations (\ref{gkgm},\ref{gkgmb}). Choosing
(\ref{gkgmb}) whose first few terms are simpler we obtain
\begin{eqnarray}
&&
g_{\mathfrak{m}}=g_{\mathfrak{m}}\left( m\right) =m+\frac{1}{2}\left[
k\left( m\right) ,m\right] +\frac{1}{12}\left[ k\left( m\right) ,\left[
k\left( m\right) ,m\right] \right] +...\text{.},\label{gm(m)}
\end{eqnarray}
which is an infinite series with rapidly decreasing coefficients. As
$g_{\mathfrak{m}}\left( m\right) $ is a converging series we can truncate
it so that the resulting equation will provide an \emph{approximation}
of $m$ with an error that decreases by including higher commutator terms.
If we call $\widetilde{P}_{p}\left( m\right) $ the truncation
that includes all terms with up to $p$ commutators, i.e.
\begin{equation}
g_{\mathfrak{m}}\left( m\right) =\widetilde{P}_{p}\left( m\right) +\text{
further terms, }
\end{equation}
\label{Pp(m)}
we can in principle solve the equation
\begin{eqnarray}
&&
P_{p}\left( m\right)
\equiv \widetilde{P}_{p}\left( m\right) -g_{\mathfrak{m}}=0
\label{Pp=0}
\end{eqnarray}
with respect to the single matrix variable $m$. However, the expression of
$k\left( m\right) $ given by (\ref{k=log(Ge-m)}) is rather complicated. So
we propose using again the BCH expansion to obtain
\begin{eqnarray}
&&
k\left( m\right) =\log (e^{g}e^{-m})=g-m-\frac{1}{2}[g,m]-\frac{1}{12}\left[
g,\left[ g,m\right] \right] +\frac{1}{12}\left[ m,\left[ m,g\right] \right]
+...\text{.} \label{k(m)=...}
\end{eqnarray}
As before, we can truncate (\ref{k(m)=...}) to a term that includes
all
$q$-th order brackets. This yields a polynomial $Q_{q}\left( m\right) $ that
approximates $k\left( m\right) $ as well as we desire at the cost of taking
extra high-order commutators
\begin{eqnarray}
&&
k\left( m\right) =Q_{q}\left( m\right) +\text{further terms.}
\label{Qq(m)}
\end{eqnarray}

After both truncations, equation (\ref{Pp=0}) is approximated by
\begin{equation*}
0=P_{p}\left( m\right) \simeq -g_{\mathfrak{m}}+m+\frac{1}{2}\left[
Q_{q}\left( m\right) ,m\right] +\frac{1}{12}\left[ Q_{q}\left( m\right) ,
\left[ Q_{q}\left( m\right) ,m\right] \right] +...\text{,}
\end{equation*}
where $P_{p}(m)$ is now a polynomial in one matrix variable with matrix
coefficients. Our problem of finding $m$ has thus been reduced to finding
the zeros of a polynomial. This can be easily performed with a numerical
algorithm \cite{Gregorio}. The accuracy of the result can be increased
by including more terms in the truncated series $P_p$ and $Q_q$,
i.e. increasing $p$ and $q$. Specifically, we prove in Appendix A
that, in this way,
the errors in determining $m$ by the truncation procedure are well
under control.

\begin{example}

Take the ``first order" truncations $p=1$, $q=1$:
\begin{eqnarray*}
P_{1}\left( m\right) &=&-g_{\mathfrak{m}}+m+\frac{1}{2}\left[ k\left(
m\right) ,m\right] , \\
Q_{1}\left( m\right) &=&g-m-\frac{1}{2}[g,m].
\end{eqnarray*}%
Then, approximating $k\left( m\right) \simeq Q_{1}\left( m\right) $, we obtain
\begin{eqnarray*}
P_{1}\left( m\right) &\simeq &-g_{\mathfrak{m}}+m+\frac{1}{2}\left[ g-m-
\frac{1}{2}[g,m],m\right] \\
&=&-g_{\mathfrak{m}}+m+\frac{1}{2}\left[ g,m\right] -\frac{1}{4}\left[ m,
\left[ m,g\right] \right].
\end{eqnarray*}
\end{example}

\subsection{\protect\bigskip The $M=K_{1}HK_{1}^{\dag }$ decomposition}

Once $m$ is known, it remains to find the subgroup element $K_{1}^{\dag }\in
\mathbf{K}$ whose adjoint action rotates $M$ onto $H=e^{h},$ $h\in
\mathfrak{ h}$.

\begin{lemma}
\label{Lema dense}The following properties are associated to $\mathbf{H}
=\exp \left( \mathfrak{h}\right) $:

\begin{enumerate}
\item $\mathbf{H}$ is a torus (compact connected Abelian Lie subgroup) of
$\mathbf{G}$; \item any vector $v\in \mathfrak{h}$ whose $1$-parameter
subgroup $\left\{ \exp \left( tv\right) \right\} $ is dense in $\mathbf{H}$
is centralized in $\mathfrak{m}$ just by $\mathfrak{h}$:
\begin{equation}
\left\{ u\in \mathfrak{m}\mid \left[ u,v\right] =0\right\} =\mathfrak{h}
\end{equation}
\label{v centralised by h}
\end{enumerate}

\begin{proof}
see e.g. \cite[\S 8.6]{Wolf}.
\end{proof}
\end{lemma}

As shown in \emph{Appendix \ref{App Cartan thm}}, the first necessary
ingredient to perform the decomposition $M=K_{1}HK_{1}^{\dag }$ is some
vector $v\in \mathfrak{h}$ that generates a dense $1$-parameter subgroup
$\exp \left( tv\right) \subset \mathfrak{h}$. This may seem abstract, but
since we have an explicit basis (\ref{recu subalgebras}) for $\mathfrak{h}$,
it suffices to take any irrational combination of the Cartan generators.

\begin{example}
In $\mathfrak{su}\left( 8\right) $, take e.g.
\begin{equation*}
v=IIX+\mathbf{\pi .}XXX+\mathbf{\pi }^{2}.YYX+\mathbf{\pi }^{3}.ZZX.
\end{equation*}
The reluctant reader may verify that, indeed, the centralizer of such $v$ in
$\mathfrak{m}$ is just $\mathfrak{h}$.
\end{example}

Now, we may define $f_{v,m}$ as in the Appendix B to be given by
\begin{eqnarray}
&&
f_{v,m}\left( K\right) =\left\langle v,\limfunc{Ad}\nolimits_{K}\left(
m\right) \right\rangle
=\dsum\limits_{a,b,c,d}C_{ad}^{c}C_{bc}^{d}v^{a}\left( K^{\dag }mK\right)
^{b}
\label{function f in detail}
\end{eqnarray}
and recover $K_{1}$ numerically as a minimum of $f_{v,m}$.

We conclude that $m$ can be rotated into $h=\limfunc{Ad} _{K_{1}}\left(
m\right) $. Thus we have completed the decomposition
\begin{equation*}
\fbox{
\begin{tabular}{l}
$\therefore G=K_{0}K_{1}e^{h}K_{1}^{\dag }$
\end{tabular}
}
\end{equation*}

\section{Step-by-step summary}
\label{Step}

What we have described so far consists of the main building blocks necessary
to perform the Khaneja-Glaser decomposition. Here, we will summarize all the
steps one needs to take when given an arbitrary unitary $G\in SU( 2^{n})$.

\begin{enumerate}
\item Calculate its (matrix) logarithm $g=\log \left( G\right) \in
\mathfrak{ g}=\mathfrak{su}\left( 2^{n}\right)$.

\item Compute the Khaneja-Glaser basis following the recurrence in diagram
$\left( \ref{Khaneja basis}\right)$; take $g_{\mathfrak{m}}$, the component
of $g$ on the subspace $\mathfrak{m}_{n}$.

\item \label{num Pp(m)}Truncate (\ref{gm(m)}) including $p$-th commutators
to get $\widetilde{P}_{p}\left( m\right) $; let $P_{p}\left( m\right) =
\widetilde{P}_{p}\left( m\right) -g_{\mathfrak{m}}$.

\item \label{num Qq(m)}Truncate $\left( \ref{k(m)=...}\right) $ including
$q$-th commutator to get $Q_{q}\left( m\right) $, as in (\ref{Qq(m)}).

\item \label{num replace}Replace $Q_{q}\left( m\right) $ for $k\left( m\right) $ in the
expression of $P_{p}\left( m\right) $ obtained in $\ref{num Pp(m)}$, so
that $P_{p}\left( m\right) $ becomes a polynomial in $m$.

\item \label{num m}Solve the zeros of $P_{p}\left( m\right) $ to get a
solution $m$ to $\left( \ref{Pp=0}\right) $.

\item \label{num K0}Use $m$ from item $\ref{num m}$ above to calculate
$K_{0}=Ge^{-m}$ as in $\left( \ref{k=log(Ge-m)}\right) $.

\item \label{num v}Compute $\mathfrak{h}_{n}$ following $\left( \ref{recu
subalgebras}\right) $; order its elements $\left\{ u_{j}\right\} $
e.g.
alphabetically and define $v=\sum \mathbf{\pi }^{j-1}u_{j}$ to satisfy the
density hypothesis of \emph{Lemma \ref{Lema dense}}.

\item \label{num K1}Use $m$ and $v$ to define $f_{v,m}\left( K\right)
=\dsum\limits_{a,b,c,d}C_{ad}^{c}C_{bc}^{d}v^{a}\left( K^{\dag }mK\right)
^{b}$ as in $\left( \ref{function f in detail}\right) $; minimize $f$ on
$\mathbf{K}=\exp \left( \mathfrak{k}_{n}\right) $ to find $K_{1}$.

\item \label{num h}Calculate $h=K_{1}^{\dag }mK_{1}$, and thus $H=\exp
\left( h\right) $.

\item \label{num partial}Assembling the results from items $\ref{num K0}$,
$ \ref{num K1}.$ and $\ref{num h}$, obtain
\begin{equation*}
G=K_{0}K_{1}HK_{1}^{\dag }.
\end{equation*}

\item \label{num for f}Repeat the above steps for $G=\widehat{K_{0}K_{1}}$
and then for $G=\widehat{K_{1}^{\dag }}$, replacing $\mathfrak{k}
_{n}\rightarrow \mathfrak{k}_{n,0}$, $\mathfrak{m}_{n}\rightarrow
\mathfrak{k }_{n,1}$ and $\mathfrak{h}_{n}\rightarrow \mathfrak{f}_{n}$.

\item Items $\ref{num partial}$ and $\ref{num for f}$ yield the
decomposition
\begin{equation*}
G=K^{\left( 1\right) }F^{\left( 1\right) }K^{\left( 2\right) }HK^{\left(
3\right) }F^{\left( 2\right) }K^{\left( 4\right) },
\end{equation*}
with $F^{\left( j\right) }\in \exp \left( \mathfrak{f}_{n}\right) ,$ $H\in
\exp \left( \mathfrak{h}_{n}\right) $ and $K^{\left( j\right) }\in SU\left(
2^{n-1}\right) \otimes SU\left( 2\right) $.

\item Decrease $n\rightarrow n-1$ and iterate this process to further
decompose each factor $K^{\left( j\right) }\in SU\left( 2^{n-1}\right)
\otimes SU\left( 2\right)$ until they all reduce to a product of Cartan factors
$F_{n}^{\left( j\right) }$ and $H_{n}^{\left( l\right) }$ and local
unitaries in $SU\left( 2\right) ^{\otimes N}$.
\end{enumerate}

NB.: As far as accuracy in step \ref{num m} is concerned, tasks
$\ref{num Pp(m)}$ to $\ref{num replace}$ should be performed in the
light of {\em Appendix \ref{Truncation}}. Namely, truncations at
higher order should be tried until numerical errors are
satisfactory, which will happen after a finite number of attempts.

\section{Conclusions}
\label{Conclusions}

As the advances in quantum technologies move beyond the control of one or
two spins, or qubits, it is important to minimize the overall cost of
processing quantum information. The Khaneja-Glaser decomposition of
$SU(2^N)$ offers an upper bound for this optimization procedure given by
$4^{N-1}$ multi-local $SU(2)^{\otimes N}$ rotations together with
$4^{N-1}-1$ purely entangling operations. The latter can be reduced,
if
desired, into bipartite interactions, or two-qubit gates. Moreover, in
\cite{Nielsen}, Nielsen gives lower bounds for such optimization. Here we
exploit the Khaneja-Glaser approach to build a constructive method for
decomposing a general unitary in terms of its local unitary components.
Abstract as it may seem, the decomposition problem can be cast in such
a way
that can be easily solved by a numerical algorithm that can be found
at
\newline
{\em http://cam.qubit.org/users/jiannis/lie\_solve[1].tar.gz}.

Finally, one should notice that the solutions we obtained are not
necessarily unique. In general, neither the zeros of the matrix polynomials,
$P_p(m)$, nor the minima of the functions, $f$, are unique. In particular,
viewing the minimization procedure from the equivalent point of view of the
diagonalization of the matrix $m$, where $K_1$ is constructed out of
the
eigenvectors of $m$, there are many equivalent solutions depending on the
particular ordering of the eigenvectors. Moreover, one should also take into
account that the exponential function has a natural $2\pi$ periodicity and
the adjoint action is $\mathbb{Z}_2$-symmetric
\cite{Kraus,SU(4)VidalCirac}. While our approach is not concerned with
the actual
parametrizations of the group elements, this is an important issue
which should be addressed in the future.

\bigskip
{\bf Acknowledgements}

We would like to thank Greg\'orio Malajovich for very fruitful
correspondence and for writing the program to find the zeros of our polynomials. 
We also thank Tim Perutz for drawing our attention to
Rouch\'e's theorem. This work was partially supported by the Royal
Society.

\newpage
\appendix

\section{Appendix: Accuracy of BCH truncations}
\label{Truncation}

Here, a generalised version of Rouch\'{e}'s theorem \cite{Lloyd} is
employed to show that the truncations performed in \emph{Subsection}
\ref{m} yield a rigorous approximation for the zeros of (\ref{gm(m)}).

\begin{theorem}
Let $\varphi ,\psi :\mathbb{C}^{r}\rightarrow \mathbb{C}^{r}$ be holomorphic
functions and $D\subset \mathbb{C}^{r}$ be an open domain such that neither $%
\varphi $ nor $\psi $ have zeros on $\partial D$; if
\begin{eqnarray}
&&
\left\vert \varphi \left( m\right) -\psi \left( m\right) \right\vert
<\left\vert \varphi \left( m\right) \right\vert +\left\vert \psi \left(
m\right) \right\vert ,\quad \forall m\in \partial D,  \label{Rouche}
\end{eqnarray}%
then $\varphi $ and $\psi $ have the same number of zeros in $D$.
\end{theorem}

NB.: We adopt, e.g., the norm $\left\vert \varphi \right\vert
=\max_{i,j}\left\vert \varphi _{ij}\right\vert $, but the argument
holds for any $L^p$-norm.

Let \ $r=2^{2n}$ be the number of entries of a matrix $m\in \mathfrak{su}%
\left( 2^{n}\right) $, seen as a complex vector. Consider then a BCH-type
series $\varphi \left( m\right) $ and its truncated version $\psi \left(
m\right) =P_{p}\left( m\right) $ containing all its terms of up to $p$
successive brackets, calling $R_{p}\left( m\right) $ the truncation
remainder
\begin{equation*}
\varphi \left( m\right) =P_{p}\left( m\right) +R_{p}\left( m\right) .
\end{equation*}%
Suppose $\hat{m}\in \mathfrak{su}\left( 2^{n}\right) $ is a zero of $%
P_{p}\left( m\right) $; then $\varphi \left( m\right) $ will also have a
zero inside the polydisc $D=\Delta _{\delta }(\hat{m})\subset
\mathbb{C}^{r}$ of radius $\delta >0$ about $\hat{m}$ if the following
(stronger) instance of $\left( \ref{Rouche}\right) $ holds
\begin{eqnarray}
&&
\left\vert R_{p}\left( m\right) \right\vert <\left\vert P_{p}\left( m\right)
\right\vert ,\quad \forall m\in \partial \Delta _{\delta }(\hat{m}).
\label{boundary}
\end{eqnarray}
In other words, $\hat{m}$ approximates at least one zero of $\varphi \left(
m\right) $ with error inferior to an arbitrarily chosen $\delta $.

All we have to show is that condition $\left( \ref{boundary}\right) $ holds
for suitably large $p$; this is a relatively straightforward consequence of
the uniform convergence of the BCH series $\varphi \left( m\right)
=\lim\limits_{p\rightarrow \infty }P_{p}\left( m\right) $, as we will now
see. For any (small) $\varepsilon _{0}>0$, there is a $p_{0}$ such that
\begin{subequations}
\begin{eqnarray}
&&
p>p_{0}\Rightarrow \left\vert P_{p}\left( m\right) -P_{p_{0}}\left( m\right)
\right\vert <\varepsilon _{0},\quad \forall m,  \label{bla0}
\end{eqnarray}
hence
\begin{eqnarray}
&&
\left\vert P_{p}\left( m\right) \right\vert >\left\vert P_{p_{0}}\left(
m\right) \right\vert -\varepsilon _{0},\quad \forall m.  \label{bla1}
\end{eqnarray}%
In particular, if $m_{0}$ is a zero of $P_{p_{0}}\left( m\right) $, equation
$\left( \ref{bla0}\right) $ restricted to the boundary $\partial \Delta
_{\delta }(m_{0})$ implies that all polynomials $P_{p}\left( m\right) $, $%
p>p_{0}$, also have at least one zero inside $\Delta _{\delta }(m_{0})$, by
Rouch\'{e}'s theorem.

On the other hand, convergence also implies $\lim\limits_{p\rightarrow
\infty }\left\vert R_{p}\left( m\right) \right\vert =0$, hence, for a given $%
\varepsilon _{1}>0$, there is $p_{1}$ such that
\end{subequations}
\begin{eqnarray}
&&
p>p_{1}\Rightarrow \left\vert R_{p}\left( m\right) \right\vert <\varepsilon
_{1},\quad \forall m.  \label{bla2}
\end{eqnarray}
Set $\varepsilon _{1}=\min\limits_{m\in \partial \Delta _{\delta
}(m_{0})}\left\vert P_{p_{0}}\left( m\right) \right\vert -\varepsilon _{0}$;
assuming this is positive (if not, take a larger $p_{0}$ for a smaller $%
\varepsilon _{0}$), take some $p>\max \left\{ p_{0},p_{1}\right\} $ and find
$m_{1}\in \Delta _{\delta }(m_{0})$ such that $P_{p}\left( m_{1}\right) =0$.
Then consider any smaller polydisc $\Delta _{\delta ^{\prime
}}(m_{1})\subset $ $\Delta _{\delta }(m_{0})$, and restrict equations $%
\left( \ref{bla1}\right) $ and $\left( \ref{bla2}\right) $ to its boundary.
We obtain%
\begin{equation*}
\left\vert R_{p}\left( m\right) \right\vert <\left\vert P_{p}\left( m\right)
\right\vert ,\quad \forall m\in \partial \Delta _{\delta ^{\prime }}(m_{1}),
\end{equation*}%
thus, by Rouch\'{e}'s theorem again, $\varphi \left( m\right) =$ $%
P_{p}\left( m\right) +R_{p}\left( m\right) $ has a zero inside $\Delta
_{\delta ^{\prime }}(m_{1})\subset $ $\Delta _{\delta }(m_{0})$.

\section{Appendix: Proof of \emph{Proposition \protect\ref{Prop Cartan
subalgebras}}\label{App Cartan thm}}

In this \emph{Appendix} we will develop the proof of \emph{Proposition
\ref{Prop Cartan subalgebras}}, following $\cite[\S 8.3]{Wolf}$. This
argument contains most of the crucial elements to understanding the
$\mathbf{KHK}$ decomposition in detail.

\begin{proof}[Proof of \emph{Proposition \protect\ref{Prop Cartan
subalgebras}}]
\begin{enumerate}
\item Given $m\in \mathfrak{m},$ we want to find $K_{1}^{\dag }\in
\mathbf{K} $ whose action rotates $m$ onto some element $h$ of the Cartan
subalgebra $\mathfrak{h}$. First, take any $v\in \mathfrak{h}$ that
generates a dense $1$ -parameter subgroup $\exp \left( tv\right) \subset
\mathbf{H}=\exp \left( \mathfrak{h}\right) $, as in \emph{Lemma \ref{Lema
dense}} and define the function
\begin{eqnarray}
f_{v,m} &=&f:\mathbf{K\rightarrow }\mathbb{R}  \label{function f} \\
K &\mapsto &f(K)=\left\langle v,\limfunc{Ad}\nolimits_{K}\left( m\right)
\right\rangle   \notag
\end{eqnarray}
where $\left\langle a,b\right\rangle =\limfunc{tr}\left( \limfunc{ad}
\nolimits_{a}\limfunc{ad}\nolimits_{b}\right) $ is the Killing form on $
\mathfrak{g}$. Since $f$ is continuous and $\mathbf{K}$ is compact, $f$
admits an absolute minimum $K_{1}\in \mathbf{K}.$ If we consider a local
perturbation of $K_{1}$ by $e^{tk}$, for any $k\in \mathfrak{k}$, we have
\begin{eqnarray*}
0 &=&\left. \frac{d}{dt}\right\vert _{t=0}f\left( e^{tk}K_{1}\right) =\left.
\frac{d}{dt}\right\vert _{t=0}\left\langle v,\limfunc{Ad}\nolimits_{e^{tk}}
\limfunc{Ad}\nolimits_{K_{1}}\left( m\right) \right\rangle  \\
&=&\left\langle v,\left[ k,\underset{h}{\underbrace{\limfunc{Ad}
\nolimits_{K_{1}}\left( m\right) }}\right] \right\rangle =\limfunc{tr}\left(
\limfunc{ad}\nolimits_{v}\limfunc{ad}\nolimits_{\left[ k,h\right] }\right)
\\
&=&\limfunc{tr}\left( \limfunc{ad}\nolimits_{v}\left( \limfunc{ad}
\nolimits_{k}\limfunc{ad}\nolimits_{h}-\limfunc{ad}\nolimits_{h}\limfunc{ad}
\nolimits_{k}\right) \right) \overset{\left( \text{c.p.t.}\right) }{=}
\limfunc{tr}\left( \limfunc{ad}\nolimits_{\left[ h,v\right] }\limfunc{ad}
\nolimits_{k}\right)  \\
&\Leftrightarrow &\left\langle \left[ h,v\right] ,k\right\rangle =0,\text{ }
\forall k\in \mathfrak{k}.
\end{eqnarray*}
But the Killing form is non-degenerate on $\mathfrak{k}$, so we must have $
\left[ h,v\right] =0$ and thus $h\in \mathfrak{h}$, as $v$ is centralized by
$\mathfrak{h}$ $\left( \ref{v centralised by h}\right) $; hence, we have
shown
that%
\begin{equation*}
m=\limfunc{Ad}\nolimits_{K_{1}^{\dag }}\left( h\right) ,\text{ }h\in
\mathfrak{h}.
\end{equation*}

\item By $1.$ there exists $K\in \mathbf{K}$ such that
\begin{equation*}
\limfunc{Ad}\nolimits_{K}\left( v\right) \in \mathfrak{h}^{\prime }.
\end{equation*}

Now take centralizers on both sides.
\end{enumerate}

\end{proof}

\end{document}